\documentstyle[12pt,epsfig,notebook]{article}
\title{\bf Wave packets and initial conditions in quantum cosmology}
\author{S. S. Gousheh and H. R. Sepangi\thanks{email: hr-sepangi@cc.sbu.ac.ir}
\\ {\small Department of Physics, Shahid Beheshti University,
Evin, Tehran 19839, Iran}}
\begin{document}
\maketitle 
\begin{abstract}
We discuss the construction of wave packets resulting from the
solutions of a class of  Wheeler-DeWitt equations in
Robertson-Walker type cosmologies. We present an ansatz for the
initial conditions which leads to a unique determination of the
expansion coefficients in the construction of the wave packets
with probability distributions which, in an interesting contrast
to some of the earlier works, agree well with all possible
classical paths. The possible relationship between these initial
conditions and signature transition in the context of classical
cosmology is also discussed. \vspace{5mm} \vspace{15mm}
\end{abstract}

\section{Introduction}
The question of the construction and interpretation of  wave
packets in quantum cosmology has been attracting much attention in
recent years. Numerous studies have been done to obtain a quantum
theory for gravity and to understand its connection with classical
physics. This interest is due to our desire for obtaining the
basic dynamical laws for predicting  the large scale structure of
the universe and the determination of the initial conditions from
which the universe has evolved. The latter is of particular
importance since if we accept that the fundamental laws of physics
are quantum mechanical in nature, then the question of initial
conditions is embodied in finding the initial condition for the
quantum cosmological state of the universe, see for example
\cite{hartle}. The search for the fundamental dynamical laws has
been underway since Newtonian times but that for the initial
conditions is relatively recent and great efforts have been made
in this direction. This is, of course, not surprising as the
underlying theories for describing the fundamental laws of physics
are local quantum field theories above the Plank length. However,
theories explaining the initial quantum state of the universe are
non-local. They imply regularities in space that have emerged on
large cosmological scales. It is only in recent years that the
progress in observational cosmology on large enough scales of both
space and time may present us with the opportunity to look at the
viability of the predictions of a theory of the initial state of
the universe.

The problem of the relation between quantum cosmology and
classical physics is an important one that exists even in simple
models \cite{pad}. Most authors consider semi-classical
approximations to the Wheeler-DeWitt (WD) equation and refer to
regions in configuration space where these solutions are
oscillatory or exponentially decaying as representing classically
allowed or forbidden regions, respectively. These regions are
mainly determined by the initial conditions for the wave function.
Two popular proposals for the initial conditions are  the {\it no
boundary proposal} \cite{hawking} and the {\it spontaneous
nucleation from nothing} \cite{vilenkin}. These proposals have
been  attractive to many authors because they lead to some classes
of classical solutions represented by certain trajectories which
posses important features such as predicting an inflationary
phase.

In quantum cosmology, in analogy with ordinary quantum mechanics,
one is generally concerned with the construction of wave functions
by the superposition of the `energy eigenstates' which would peak
around the classical trajectories in configuration space. However,
contrary to ordinary quantum mechanics, a parameter describing
time is absent in quantum cosmology so that the initial conditions
would have to be expressed in terms of an {\it intrinsic} time
parameter, which in the case of the WD equation could be taken as
the local scale factor for the three geometry \cite{dewitt}. Also,
since the sign of the kinetic term for the scale factor is
negative, a formulation of the Cauchy problem for the WD equation
is possible. The existence of such a sign is one of the exclusive
features of gravity with many other interesting implications.

The construction of wave packets resulting from the solutions of
the WD equation has been a feature common to most of the recent
research work in quantum cosmology \cite{kiefer,tucker,sepangi}.
In particular, in reference \cite{kiefer} the construction of wave
packets in a Friedmann universe is presented in detail and
appropriate boundary conditions are motivated. Generally speaking
one of the aims of these investigations has been to find wave
packets whose probability distributions coincide with the
classical paths obtained in classical cosmology. In these works,
the authors usually consider model theories in which a self
interacting scalar field is coupled to gravity in a
Robertson-Walker type universe. The resulting WD equation is often
in the form of an isotropic oscillator-ghost-oscillator and is
separable in the configuration space variables. The general
solution can thus be written as a sum over the product of simple
harmonic oscillator wave functions with different frequencies. As
usual, the coefficients in the sum are chosen according to the
initial conditions, which are usually specified by giving the wave
function at $R=0$, where $R$ is the scale factor. The choice of
the initial conditions, or equivalently the coefficients in the
sum, varies considerably amongst various authors. In
\cite{tucker,sepangi}, the wave packets are constructed by summing
over the first few terms with the coefficients being chosen
arbitrarily to be equal. However, in \cite{kiefer} the initial
state is chosen to be a Gaussian with the appropriate symmetry and
the  undetermined coefficients being set to zero. The resulting
wave packets, in general, lack one or more of the following
desired properties. Firstly, one expects the wave packet
describing a physical system to posses a certain degree of
smoothness. Secondly, there should be a good classical-quantum
correspondence, which means that not only the wave packet should
be centered around the classical path, but also the crest of the
probability distribution should coincide as closely as possible
with the classical trajectory. Also, to each distinct classical
path there should correspond a unique wave packet.

The purpose of this paper is to address and clarify some of these issues.
We first argue that specifying
the wave function at $R=0$ does not furnish a complete set of initial
conditions.
We then
suggest
an ansatz for the expansion coefficients, or equivalently for the initial
conditions, which produces wave packets with the desired properties mentioned
above. The organization of the paper is a follows: In section two we give a
short
review of some of the models that lead to the class of WD equations considered
here.
Section three deals with
the construction of the wave packets and in section four we discuss
the salient features of this work and conclusions are drawn.

\section{Robertson-Walker cosmology with coupled scalar field}

In Robertson-Walker cosmology, one often considers the coupling
of a scalar field  to gravity. The
resulting field equations include a `zero energy constraint'.
The WD equation in quantum cosmology is the result of the quantization
consistent with this constraint.
In some simple minisuperspace models the solutions to the WD equation describe
a system of oscillators with unequal or equal frequencies.
It would therefore be
appropriate at this point to give a brief review of some of the minisuperspace
models
leading to the class of WD equations whose solutions are to be discussed here.

Consider Einstein's field equation coupled to a scalar field
\begin{eqnarray}
R_{\mu\nu}-\frac{1}{2}g_{\mu\nu}R&=&\kappa T_{\mu\nu}(\phi), \label{eq1} \\
\Delta^2\phi-\frac{\partial U}{\partial \phi}&=&0, \label{eq2}
\end{eqnarray}
where $T_{\mu\nu}$ is the energy-momentum tensor coupling the
scalar field $\phi$ to gravity and $U$ is a scalar potential for
the scalar field $\phi$. Hereafter, we shall work in a system of
units in which $\kappa=1$. We parameterize the metric as
\begin{eqnarray}
g=-dt^2+R^2(t)\frac{\sum dr^i dr^i}{(1+kr^2/4)^2},  \label{eq3}
\end{eqnarray}
where $R(t)$ is the usual scale factor and $k=+1,0,-1$ corresponds
to a closed, flat or open universe respectively. The field equations
resulting from
(\ref{eq1}) and (\ref{eq2}) with the metric given by (\ref{eq3}) can be
written as
\begin{eqnarray}
3\left[\left(\frac{\dot{R}}{R}\right)^2+\frac{k}{R^2}\right]&=&
\frac{\dot{\phi}^2}{2}+U(\phi), \label{eq4}\\
2\left(\frac{\ddot{R}}{R}\right)+\left(\frac{\dot{R}}{R}\right)^2+
\frac{k}{R^2}&=&-\frac{\dot{\phi}^2}{2}+U(\phi), \label{eq5}\\
\ddot{\phi}+3\frac{\dot{R}}{R}\dot{\phi}+\frac{\partial U}{\partial\phi}
&=&0.  \label{eq6}
\end{eqnarray}
Here a dot represents differentiation with respect to time. The choice of
$U(\phi )$ is important and a particularly interesting choice with three
free parameters is \cite{tucker}
\begin{eqnarray}
U(\phi)=\lambda+\frac{m^2}{2\alpha^2}\sinh^2(\alpha\phi)+
\frac{b}{2\alpha^2}\sinh(2\alpha\phi).  \label{eq7}
\end{eqnarray}
In the above expression $\lambda$ may be identified with a cosmological
constant,
$m^2=\partial^2 U/\partial\phi^2|_{\phi=0}$ is a mass squared parameter,
$b$ is a coupling constant and $\alpha^2=\frac{3}{8}$.
The Lagrangian giving the above equations of motion can be written as
\begin{eqnarray}
L=-3R\dot{R}^2+3kR+R^3 [\dot{\phi}^2/2-U(\phi)].
\label{eq8}
\end{eqnarray}
This Lagrangian
can be cast into a simple form for $k=0$ describing an oscillator-ghost-
oscillator system \cite{tucker,tucker1}. Briefly, this is achieved  by using
the transformations
$X=R^{3/2} \cosh(\alpha\phi)$ and $Y=R^{3/2} \sinh(\alpha\phi)$, which
transform the term $R^3 U(\phi )$ into a quadratic form. Upon using a second
transformation to eliminate the coupling term in the quadratic form, we
arrive at new variables $u$ and $v$, which are linear combinations of $X$
and $Y$, in terms of which the Lagrangian takes on the simple form
\begin{eqnarray}
L(u,v)=\frac{1}{2}\left[ (\dot{u}^2-\omega_1^2 u^2)-(\dot{v}^2-\omega_2^2 v^2)
\right],\label{eq9}
\end{eqnarray}
where $\omega_{1,2}^2 = -3\lambda /4 +m^2/2 \mp\sqrt{m^4-4b^2}/2$.
The classical solutions for $k\ne 0$ are given in \cite{siamak}.
The corresponding quantum cosmology for $k=0$ is described by the
WD equation written as
\begin{eqnarray}
H\psi(u,v)=\left\{-\frac{\partial^2}{\partial u^2}+\frac{\partial^2}
{\partial v^2}+
\omega_1^2 u^2-\omega_2^2 v^2\right\}\psi(u,v)=0.  \label{eq10}
\end{eqnarray}

Alternatively, one may obtain a minisuperspace model described by
the same WD equation but with equal frequencies by considering a
closed Robertson-Walker universe with a vanishing cosmological
constant and containing a conformally coupled scalar field
\cite{kiefer}. A different model leading to the same WD equation
with equal frequencies is obtained by considering a Kaluza-Klein
cosmology with a negative cosmological constant described by the
metric
\begin{eqnarray}
g=-dt^2+R^2(t)\frac{\sum dr^i dr^i}{(1+kr^2/4)^2}+a^2(t)d\rho^2,
\label{eq12}
\end{eqnarray}
where $a(t)$ is the radius of the compactified space \cite{sepangi,wudka}.
The Lagrangian
describing the above cosmology is
\begin{eqnarray}
L=-\frac{1}{2}Ra\dot{R}^2-\frac{1}{2}R^2\dot{R}\dot{a}+\frac{1}{2}
kRa-\frac{1}{6}\Lambda R^3 a, \label{eq13}
\end{eqnarray}
where $\Lambda$ is the cosmological constant. By defining $\omega^2=
-\frac{2}{3}\Lambda$ and changing the variables as
\begin{eqnarray}
u=\frac{1}{\sqrt{8}}\left[R^2-Ra-\frac{3k}{\Lambda}\right], \hspace{10mm}
v=\frac{1}{\sqrt{8}}\left[R^2+Ra-\frac{3k}{\Lambda}\right], \label{eq14}
\end{eqnarray}
$L$ takes on the form
\begin{eqnarray}
L=\frac{1}{2}\left[(\dot{u}^2-\omega^2 u^2)-(\dot{v}^2-\omega^2 v^2)
\right]. \label{eq15}
\end{eqnarray}
It is easily seen that the corresponding quantum cosmology is the same as
(\ref{eq10}) with equal frequencies. We therefore focus attention on
solutions to equation (\ref{eq10}).
\section{Wave packets}
Equation (\ref{eq10}) is separable in the minisuperspace
variables and a solution can be written as
\begin{eqnarray}
\psi_{n_1 ,n_2}(u,v)=\alpha_{n_1}(u)\beta_{n_2}(v),  \label{n1n2}
\end{eqnarray}
where
\begin{eqnarray}
\alpha_n(u)=\left(\frac{\omega_1}{\pi}\right)^{1/4}\frac{H_n(
\sqrt{\omega_1}u)} {\sqrt{2^n n!}}e^{-\omega_1 u^2/2},
\label{eq16}\\
\beta_n(v)=\left(\frac{\omega_2}{\pi}\right)^{1/4}\frac{H_n(\sqrt{\omega_2}v)
} {\sqrt{2^n n!}}e^{-\omega_2 v^2/2}. \label{eq17}
\end{eqnarray}
Here $H_n(x)$ are Hermite polynomials, and the  zero energy condition leads to
\begin{eqnarray}
\left( n_1+\frac{1}{2}\right)\omega_1=\left( n_2+\frac{1}{2}\right)\omega_2,
\hspace{7mm} n_1,n_2=0,1,2,\ldots .  \label{eq18}
\end{eqnarray}
The set $\left\{ \psi_{n_1 ,n_2}(u,v) \right\} $ span the zero
sector subspace of the Hilbert space of ${\cal L}^2$ of measurable
square integrable functions on $\bf R^2$ with an inner product
defined in the usual way giving $$ \int \psi_{n_1,n_2}(u,v)
\psi_{n_1^\prime ,n_2^\prime}(u,v) du dv =\delta_{n_1,n_1^\prime}
\delta_{n_2,n_2^\prime}. $$
That is, the orthonormality and completeness of the basis
functions follow from those of the Hermite polynomials.

We can construct a general wave packet as follows
\begin{eqnarray}
\psi(u,v)={\sum_{n_1,n_2}}^\prime A_{n_1,n_2} \alpha_{n_1}(u) \beta_{n_2}(v),
\label{eq19}
\end{eqnarray}
where the prime on the sum indicates summing over all values of $n_1$ and
$n_2$ satisfying
the constraint (\ref{eq18}). As the signs of the kinetic terms in
equation (\ref{eq10}) indicate, we can
take $v$ as playing the role of the scale factor and hence
the initial condition on $\psi$ is specified by
\begin{equation}
\psi(u,0)={\sum_{n_1}}^{\prime\prime} c_{n_1}\alpha_{n_1}(u), \label{eq20}
\end{equation}
where the coefficients $c_{n_1}$ are arbitrary and the double prime on the sum
indicates that the sum runs only over
those values of $n_1$ satisfying the
constraint (\ref{eq18}) with $n_2$ being even. The constraint on the evenness
of $n_2$ is due to the fact that $\beta_{n_2}(0) =0$ for odd values of $n_2$.
We choose the coefficients $c_n$ to be the same as those of the coherent
states of a
one dimensional simple harmonic oscillator, that is
\begin{equation}
c_n=e^{-\frac{1}{4}| \chi_0 |^2}\frac{\chi_0^n}{\sqrt{2^n n!}}, \label{eq21}
\end{equation}
where $\chi_0$ is an arbitrary complex number.
Comparing the coefficients in equations (\ref{eq19},\ref{eq20}) one finds for
even values of $n_2$
\begin{eqnarray}
\left(\frac{\omega_2}{\pi}\right)^{1/4}\frac{A_{n_1,n_2}}
{\sqrt{2^{n_2} n_2!}} & = & \frac{c_{n_1}}{H_{n_2}(0)}\label{eq22a} \\
       & = & \frac{(n_2/2)!c_{n_1}}{(-1)^{n_2/2}n_2!},
       \label{eq22b}
\end{eqnarray}
and $A_{n_1,n_2}$ are arbitrary for odd values of $n_2$. This arbitrariness
is a consequence of not having specified $\left. \partial\psi (u,v)/\partial v
\right|_{v=0}$.
The WD equation is a second order PDE, and to have a unique solution one has
to specify both the wave function and its derivative at a given point.
In the first reference in \cite{kiefer} the author has chosen $A_{n_1,n_2}=0$
for odd values of $n_2$, which is equivalent to choosing
 $\left. \partial\psi (u,v)/\partial v \right|_{v=0}=0$.
This might seem to be the only choice when considering equation (\ref{eq22a}).
Here we  argue that this is not the best or the `canonical' choice.
We have chosen $A_{n_1,n_2}$ to be nonzero and to have the same functional form
for both even and odd values of $n_2$. This can be easily done using
equation (\ref{eq22b}).

The classical paths corresponding to these solutions are the generalized
Lissajous ellipsis which have the following parametric representation
\begin{eqnarray}
u(t) = u_0 \cos(\omega_1 t - \theta_0), \hspace{7mm}
v(t) = v_0 \sin(\omega_2 t) \label{eq23},
\end{eqnarray}
where the zero energy condition demands $\omega_1 u_0 = \omega_2 v_0$, and
$\theta_0$ is an arbitrary phase factor. The classical-quantum correspondence
is established by the following equation
\begin{eqnarray}
\chi_0=\sqrt{\frac{\omega_1}{\omega_2}} u_0 e^{i \theta_0}. \label{eq23.2}
\end{eqnarray}

Below we will consider several illustrative examples in order to compare our
results with some of the previous works.
Let us first consider the simplest case which is when $\omega_1=\omega_2=
\omega$. Upon inspection of equations (\ref{eq16}) through (\ref{eq19}), we
immediately recognize that,
since $n_1=n_2=n$ and $H_n(0)=0$ for odd values of $n$, firstly $A_{n,n}$
are not determined for odd values of $n$, and secondly
\begin{eqnarray}
\psi(u,0) & = & \psi(-u,0), \label{eq24}\\ \left.
\frac{\partial}{\partial v}\psi(u,v) \right|_{v=0}& = & - \left.
\frac{\partial}{\partial v}\psi(-u,v) \right|_{v=0}\label{eq25}.
\end{eqnarray}
Therefore the initial state has to be symmetric and we choose it to be
two symmetric Gaussians. However this is automatically taken into account by
the restrictions imposed on the sum in equation (\ref{eq20}) and the choice
of coefficients given in equation (\ref{eq21}). Also note that
equation (\ref{eq25}) is redundant due to the properties of the Hermite
polynomials.

Figure 1 Shows the square of the wave packet $| \psi(u,v)|^2$ for
$|\chi_0| =12$ and $\theta_0 =0$, summing over
both even and odd values of $n$, and the contour plot of the same  wave
packet along with the classical path superimposed on it.
In practice, in order to obtain a reasonable graphical representation, the
minimum number of terms included in the sum, denoted by $n_{max}$, has to be
of the order of $ | \chi_0 |^2 $. As is apparent from
these figures, the crest of the wave packet follows the classical path exactly.
This is to be compared with figure 1 of the first reference in \cite{kiefer}
which we
reproduce here as figure 2 for the same value of $n_{max}$ as in figure 1.
\begin{figure}
\centerline{\begin{tabular}{ccc}
\epsfig{figure=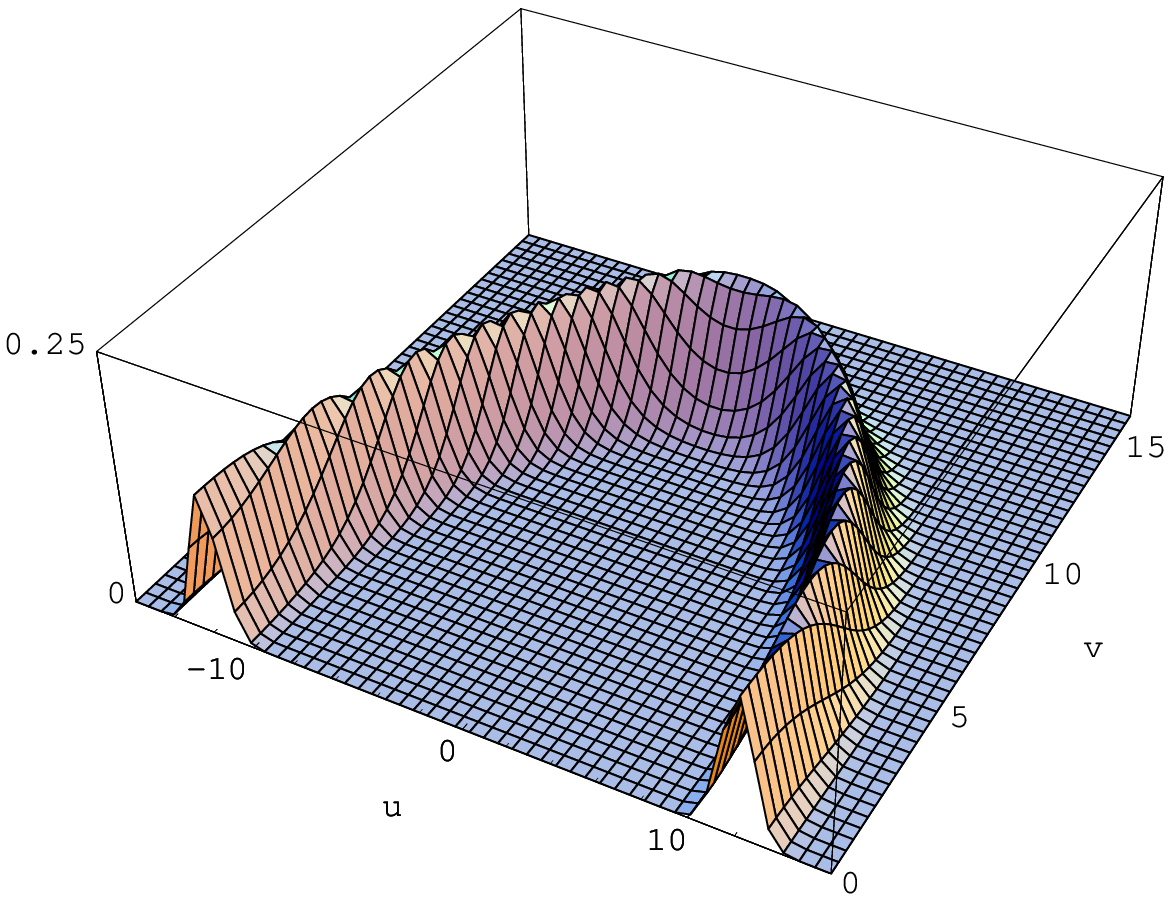,width=8cm}
 &\hspace{2.cm}&
\epsfig{figure=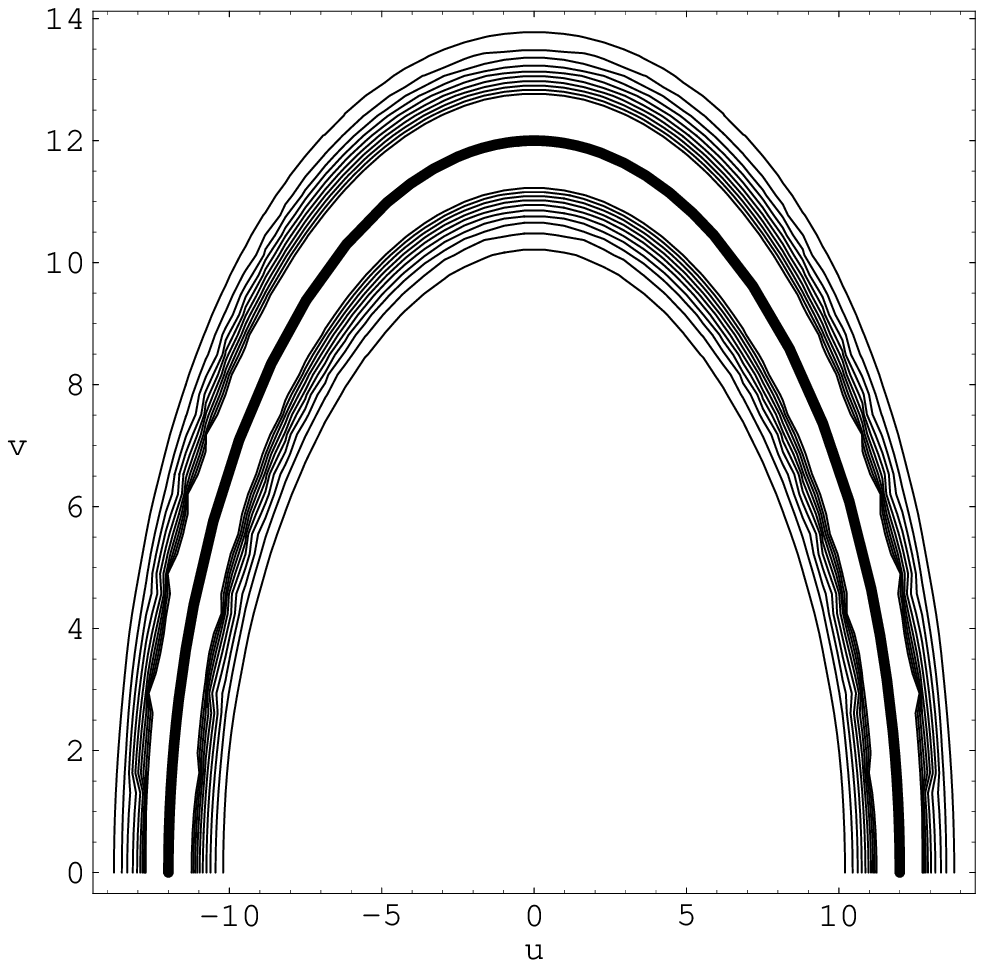,width=6.5cm}
\end{tabular}  }
\caption{Left,
the square of the wave packet $| \psi(u,v)|^2$ for $|\chi_0|=12$ and
$\theta_0 =0$, $\omega_1 =\omega_2 =1$
with $n_{max}=130$ and right, the contour plot of the same figure with the
classical path superimposed as the thick solid line.}
\label{fig1}
\end{figure}
\begin{figure}
\centerline{\begin{tabular}{ccc}
\epsfig{figure=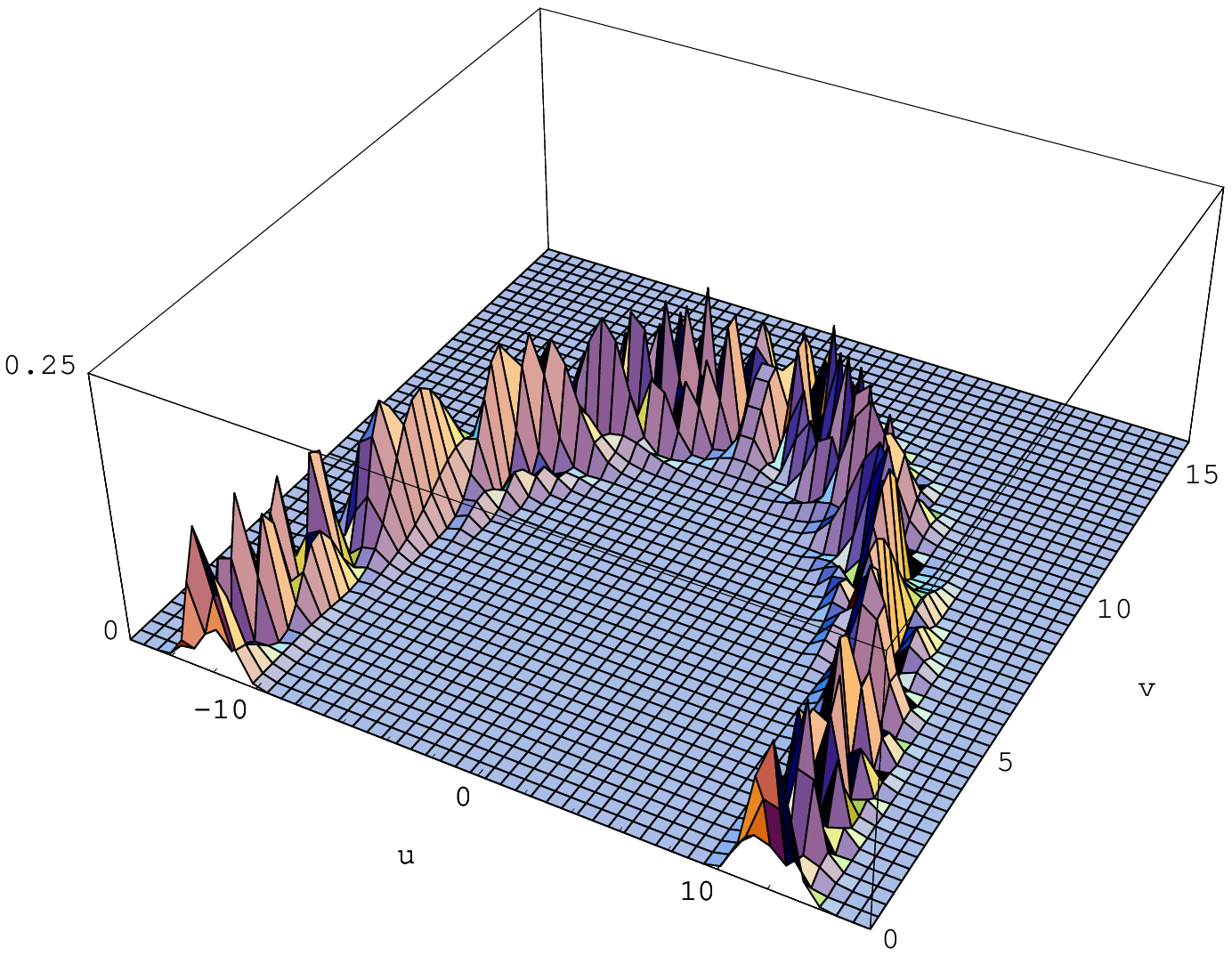,width=8.5cm}&
\hspace{2.cm}&
\epsfig{figure=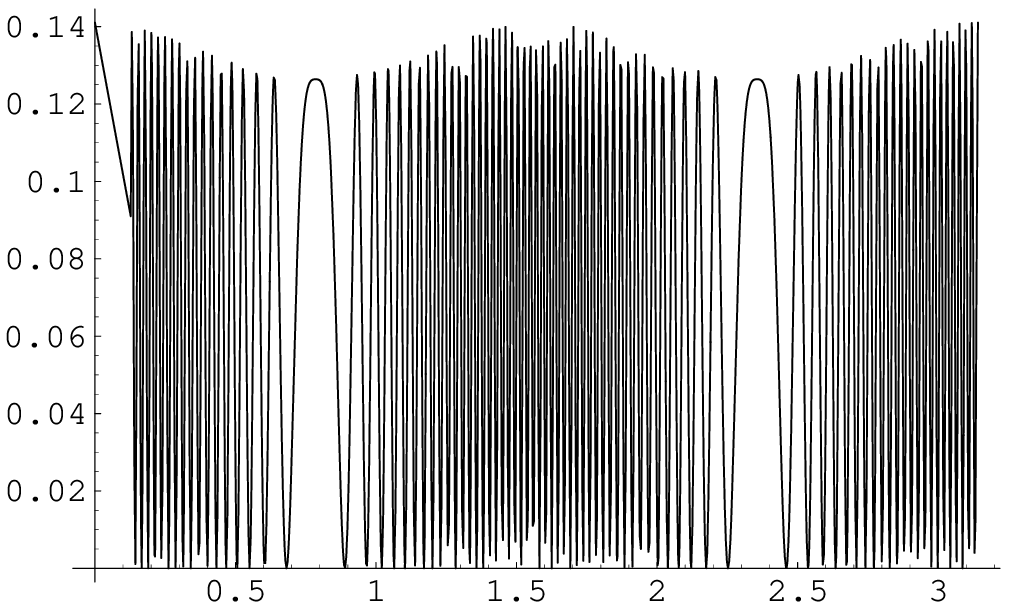,width=6.5cm}
\end{tabular} }
\caption{Left, the square of the wave packet
$| \psi(u,v)|^2$ for $|\chi_0|=12$ and $\theta_0 =0$, $\omega_1 =\omega_2 =1$
with $n_{max}=130$, summing only over even values of $n$ and right, the
parametric plot of $| \psi(u,v)|^2$ along its crest showing rapid oscillations.
The average frequency of these oscillations increases as $n_{max}$ increases.}
\label{fig2}
\end{figure}
At this stage, due to the symmetry of the circle, the only advantage of our
choice of the coefficients seems to be that the resulting wave packet is
smoother. There is another advantage which will show up in the asymmetric case
considered below.

One can easily interpolate between a circle and a
line segment simply by changing the value of $\theta_0$ between zero and $\pm
\pi /2$. Figure 3 shows the wave packet for $|\chi_0| =12$ and $\theta_0 =1$.
\begin{figure}
\centerline{\begin{tabular}{ccc}
\epsfig{figure=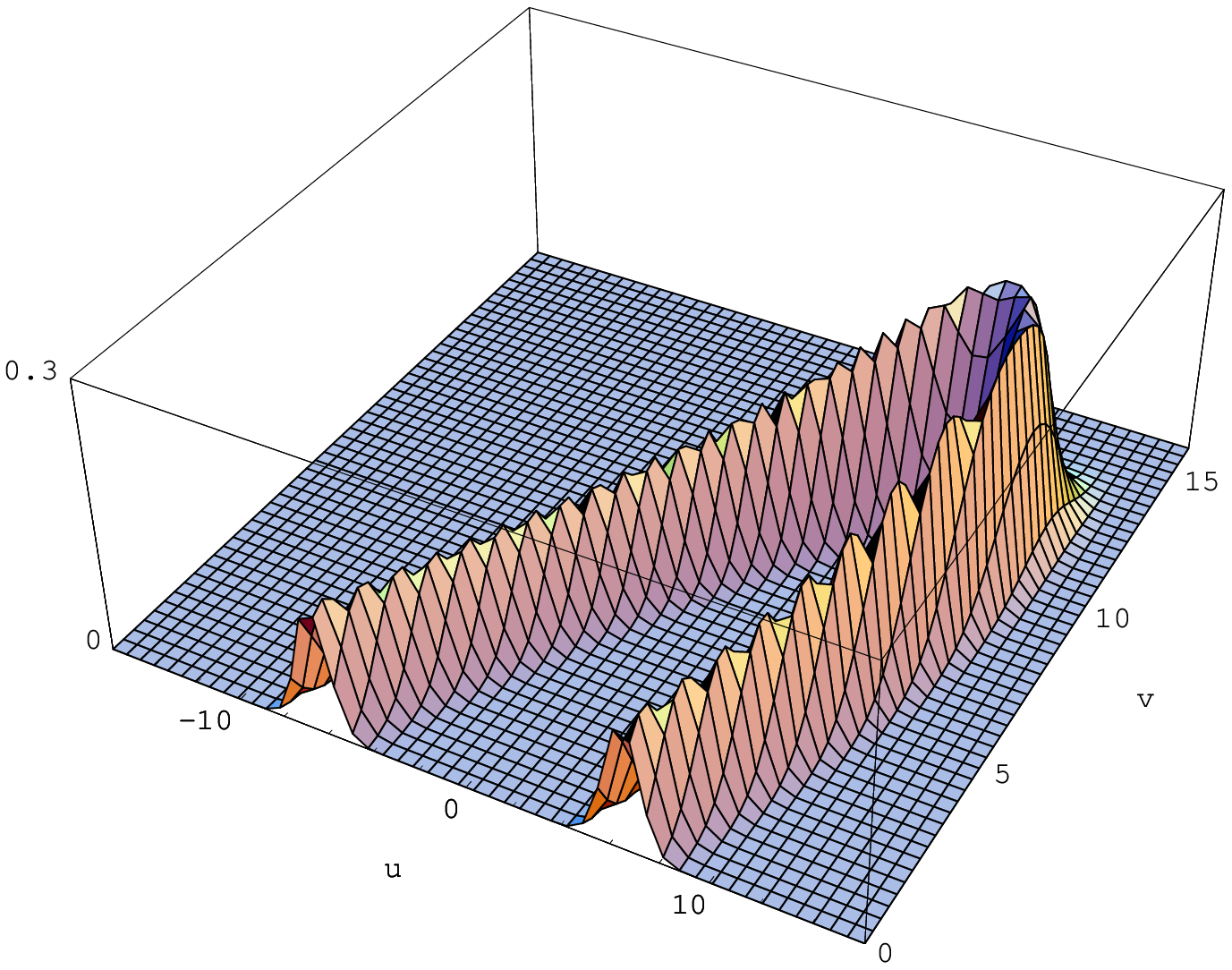,width=8cm}
 &\hspace{2.cm}&
\epsfig{figure=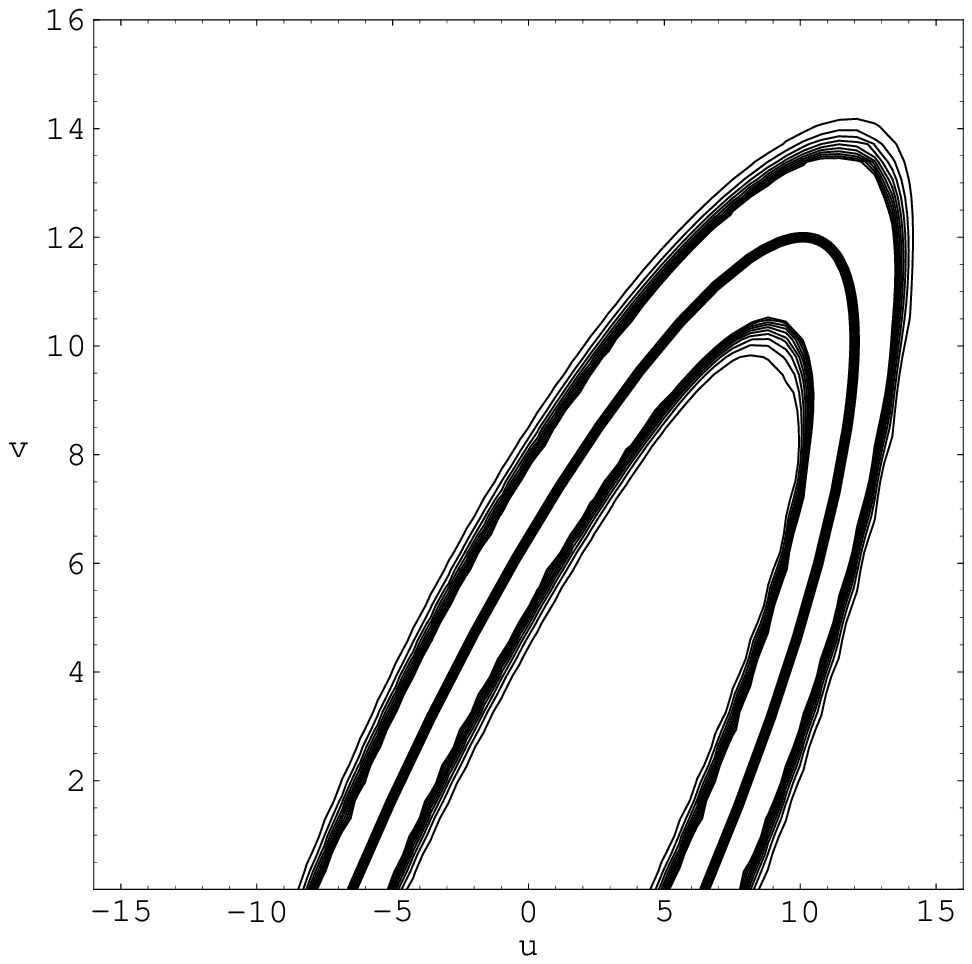,width=6.5cm}
\end{tabular}  }
\caption{Left, the square of the wave packet $| \psi(u,v)|^2$ for
$|\chi_0| =12$ and $\theta_0 =1$, $\omega_1 =\omega_2 =1$
with $n_{max}=130$ and right, the contour plot of the same figure with the
classical path superimposed as the thick solid line.}
\label{fig3}
\end{figure}
This is to be compared with figure 3 of the first reference in
\cite{kiefer}, where both wave packets with $\theta_0 =\pm 1$ are
superimposed due to the choice $A_{n,n}=0$ for odd values of $n$.
This is not a desired result since in order to establish a
classical-quantum correspondence we need to  have a separate wave
packet for each distinct classical path. At the classical level, a
given set of initial conditions, {\it e.g.} specifying
$u(0),v(0)$, uniquely determines a classical path. At the quantum
level the initial conditions are specified for example by giving
$\psi(u,0)$ and $\left. \partial \psi (u,v) / \partial v
\right|_{v=0}$, and this will uniquely determine a wave packet. Of
course the initial conditions have to be consistent with the
symmetries derived from the general structure of the solutions
given by equation (\ref{eq19}). In the present context, the first
condition determines $A_{n,n}$ for even values of $n$, and the
second the odd coefficients. Therefore the simultaneous appearance
of two wave packets corresponding to two different classical paths
is solely due to choosing $\left. \partial \psi (u,v) / \partial v
\right|_{v=0} =0$ or equivalently setting $A_{n,n}=0$ for odd
values of $n$. This is in contrast to the interpretation given by
the author in \cite{kiefer} where he relates the simultaneous
appearance of two distinct branches to the superposition
principle, and their only means of separation to the decoherence
mechanism. As we have shown, for the class of problems discussed
here, the use of the ansatz for the initial conditions given above
would result in the selection of a unique branch without resorting
to the decoherence mechanism.

As $\theta_0 \rightarrow
\pm \pi /2$, the semi-minor axis of the ellipse goes to zero. The results are
shown in figure 4. This configuration is to be compared with the
approximate one obtained by adding a few terms with equal
coefficients in \cite{sepangi}.
\begin{figure}
\centerline{\begin{tabular}{ccc}
\epsfig{figure=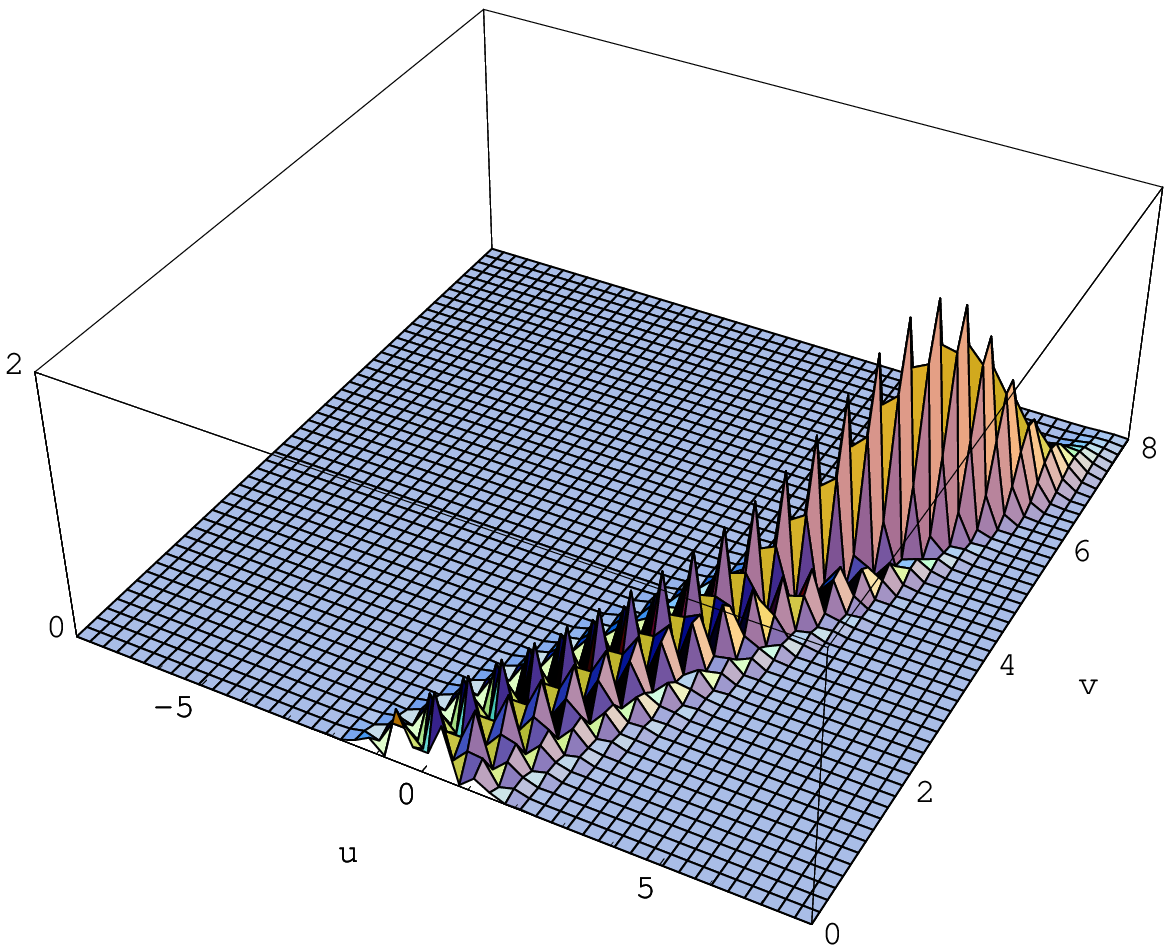,width=8cm}
 &\hspace{2.cm}&
\epsfig{figure=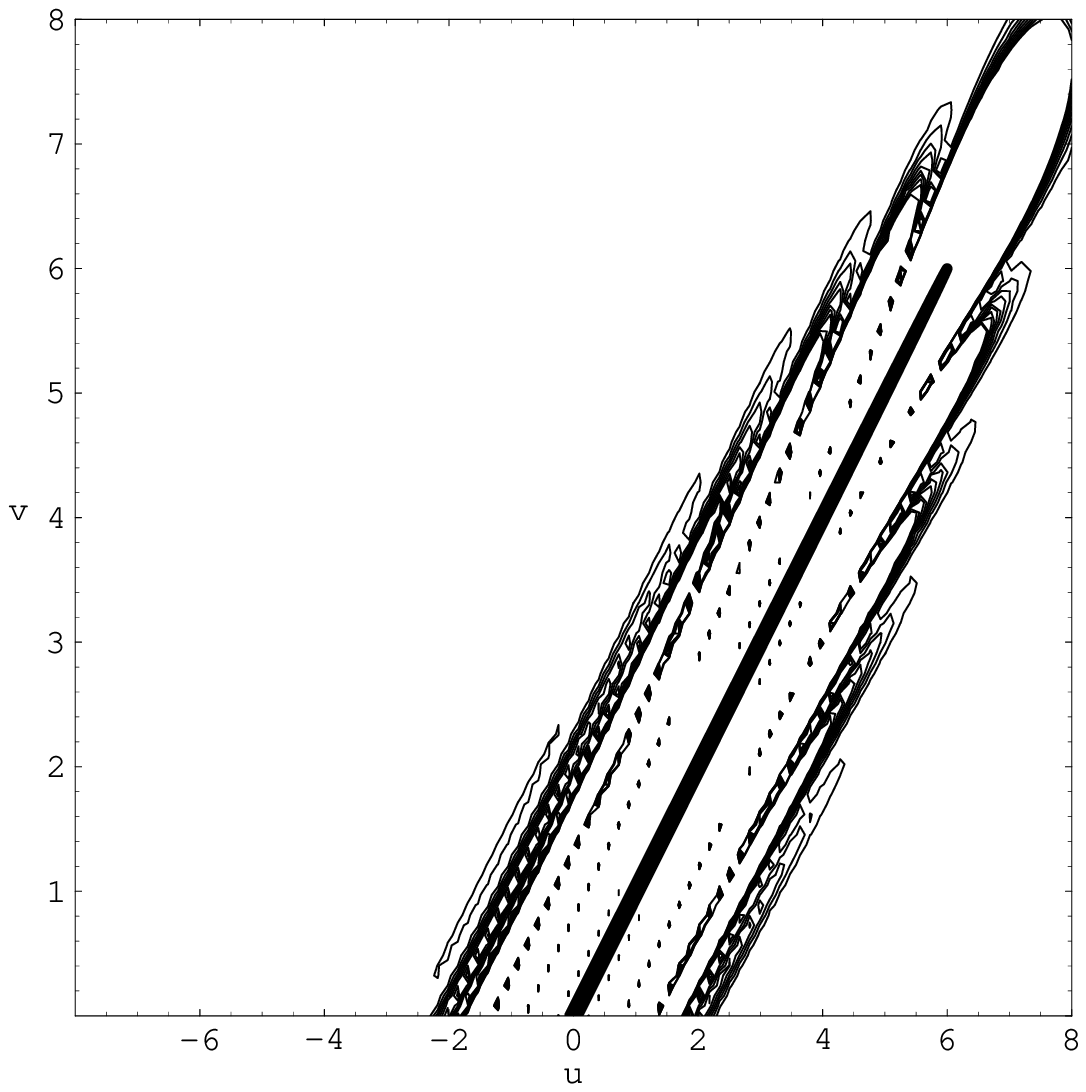,width=6.5cm}
\end{tabular}  }
\caption{Left, the square of the wave packet $| \psi(u,v)|^2$ for
$|\chi_0| =6$ and $\theta_0 = \pi $, $\omega_1 =\omega_2 =1$ with
$n_{max}=50$ and right, the contour plot of the same figure with
the classical path superimposed as the thick solid line.}
\label{fig4}
\end{figure}

Now we discuss the case with unequal frequency and to be concrete we
give an example with $\omega_1 =3 \omega_2 $. Then the constraint equation
(\ref{eq18}) gives $n_2 =3 n_1 +1$. Therefore at $v=0$ only the odd values
of $n_1$ contribute. Hence we have
\begin{eqnarray}
\psi(u,0) & = & - \psi(-u,0), \label{eq26}\\ \left.
\frac{\partial}{\partial v}\psi(u,v) \right|_{v=0}& = & \left.
\frac{\partial}{\partial v}\psi(-u,v) \right|_{v=0}\label{eq27}.
\end{eqnarray}
Again equation (\ref{eq27}) is redundant and equation (\ref{eq26}) is
automatically taken into account by the restrictions imposed on
the sum in equation (\ref{eq20}). This together with the choice of the
coefficients
given in equation (\ref{eq21}) gives two antisymmetric Gaussians.
Only the coefficients $A_{n_1,n_2}$ for odd values of $n_1$ are determined
and those for even values of $n_1$ are fixed by the prescription given above.
Figure 5 shows a plot of $ | \psi (u,v)|^2$ for $|\chi_0|=6$ and
$\theta_0 =0$. There are pronounced oscillations where there is
an overlap between different branches of the wave packet due to the
interference effect. This can be
clearly seen from a parametric plot of $ | \psi (u,v)|^2$ along the classical
path similar to figure 2, which we do not show here. In figure 5 these
oscillations are smoothed out and show up as the bump in the center of the
figure. Also note that there is a good classical-quantum correspondence.
\begin{figure}
\centerline{\begin{tabular}{ccc}
\epsfig{figure=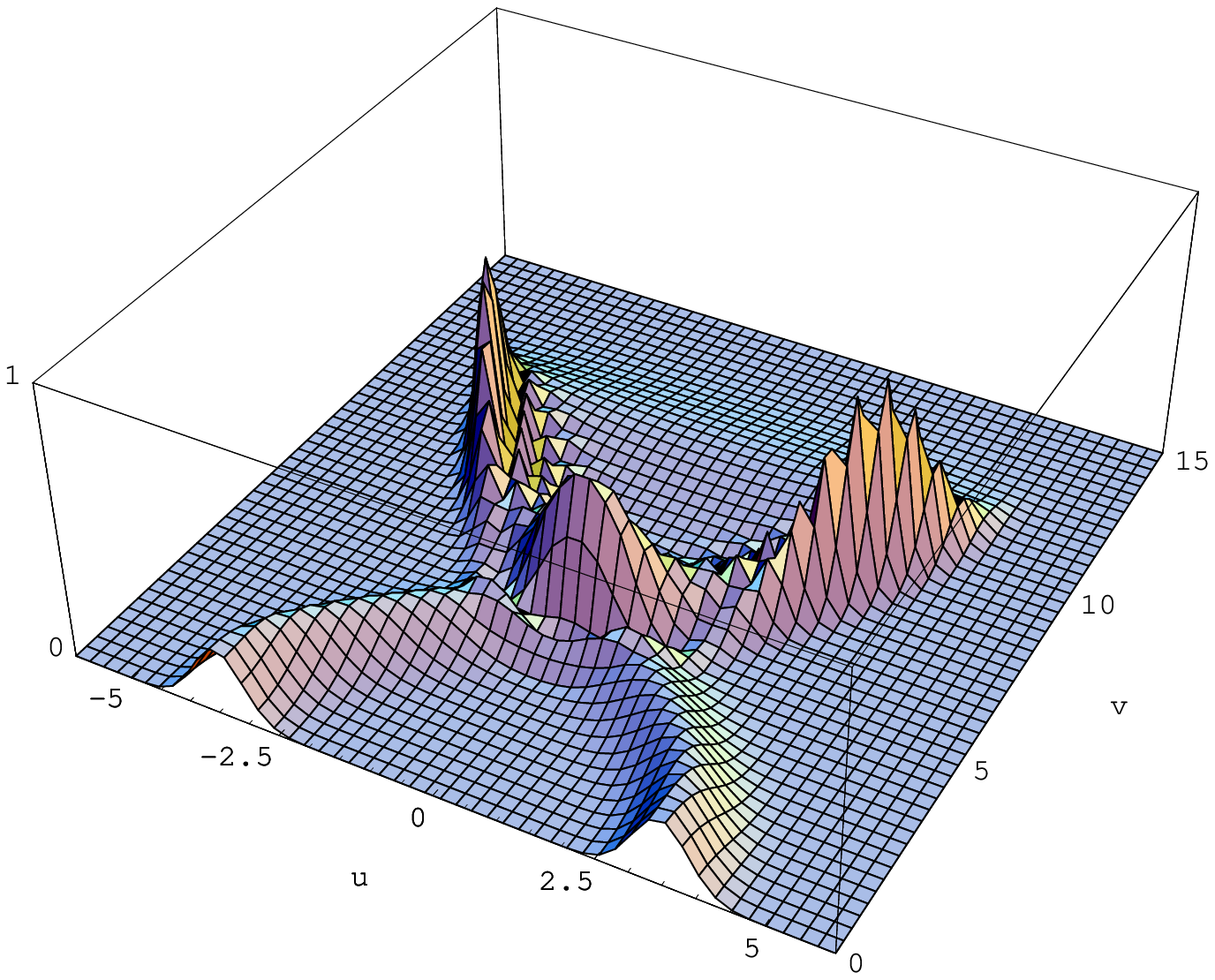,width=8cm}
 &\hspace{2.cm}&
\epsfig{figure=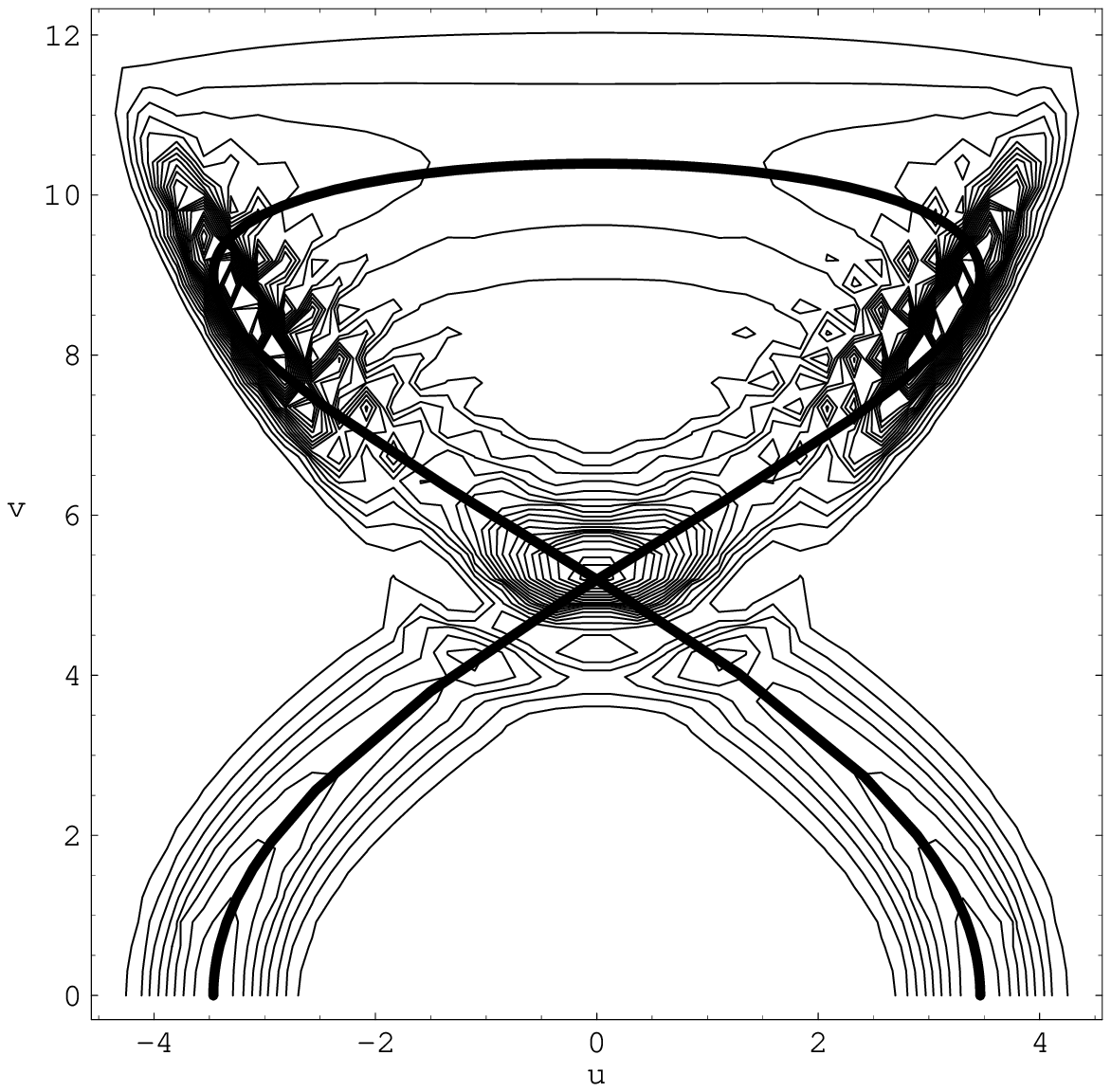,width=6.5cm}
\end{tabular}  }
\caption{Left, the square of the wave packet $| \psi(u,v)|^2$ for
$|\chi_0| =6$ and $\theta_0 =0$, $\omega_1 =3 \omega_2 =3$ with $n_{max}=50$
and right, the
contour plot of the same figure with the classical path superimposed
as the thick solid line.}
\label{fig5}
\end{figure}

Let us next consider
the case $\theta_0 =\pi /2$ whose wave packet is shown in figure 6.
\begin{figure}
\centerline{\begin{tabular}{ccc}
\epsfig{figure=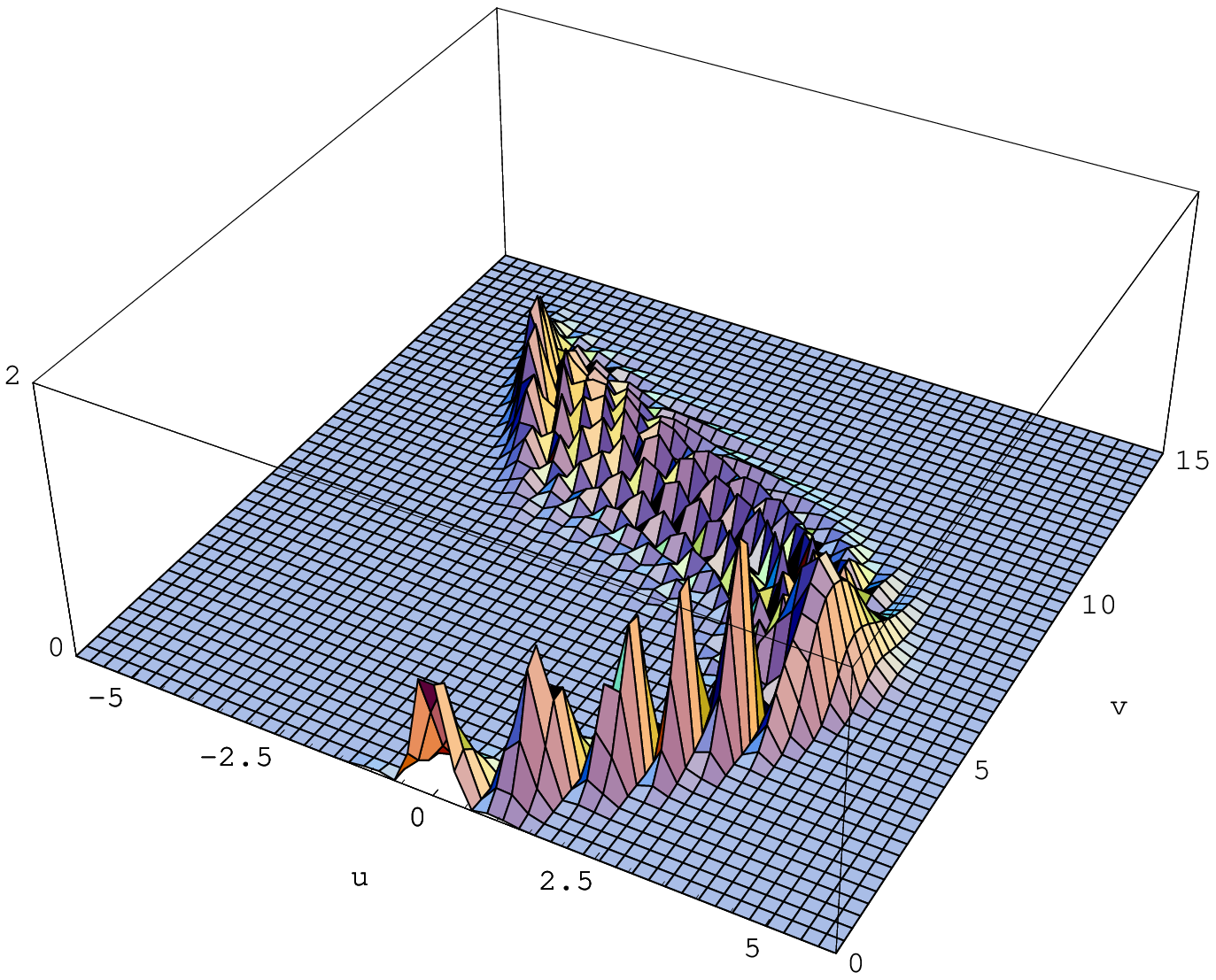,width=8cm}
 &\hspace{2.cm}&
\epsfig{figure=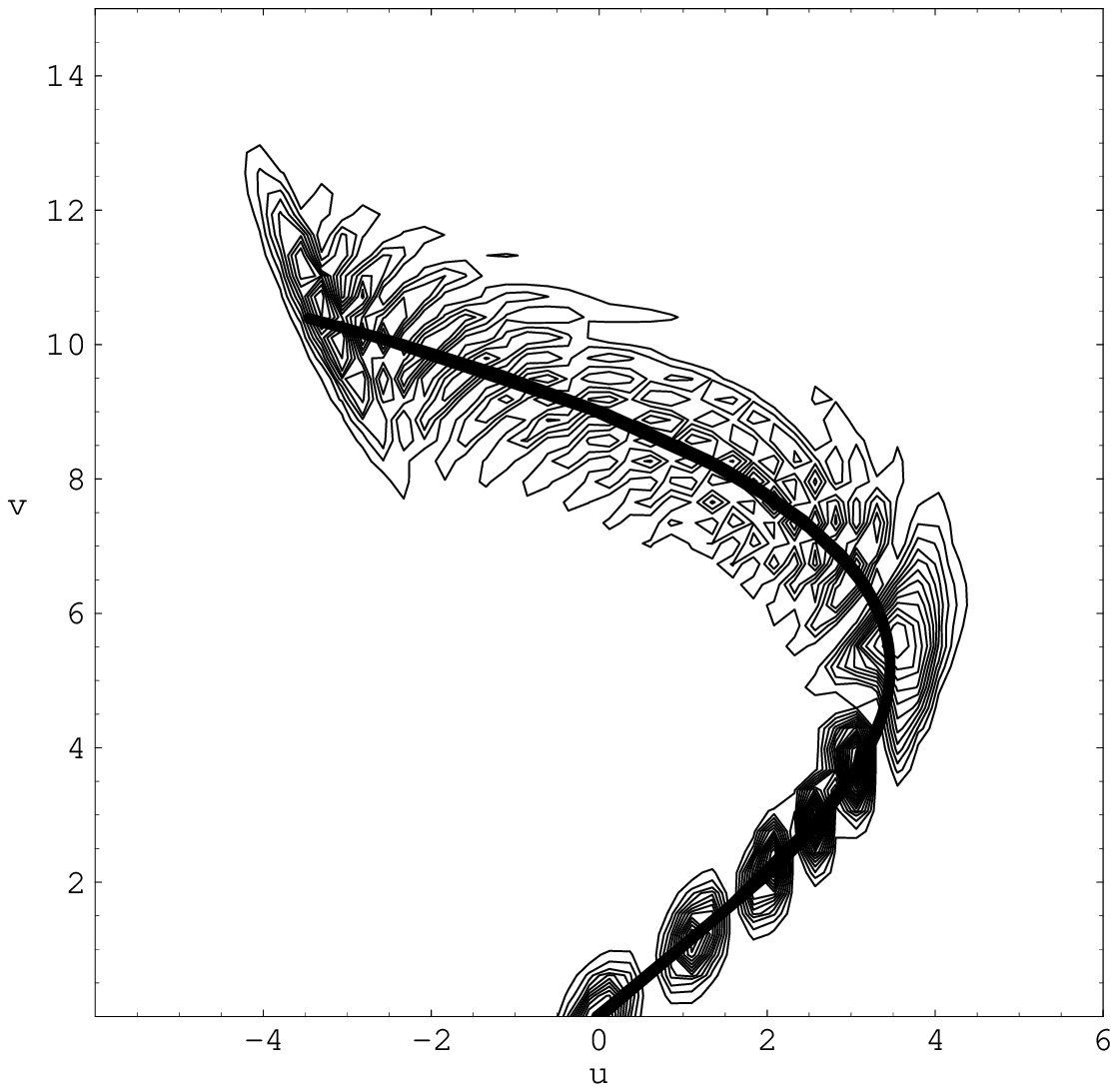,width=6.5cm}
\end{tabular}  }
\caption{Left, the square of the wave packet $| \psi(u,v)|^2$ for
$|\chi_0| =6$ and $\theta_0 =\pi $, $\omega_1 =3 \omega_2 =3$ with
$n_{max}=50$ and right, the
contour plot of the same figure with the classical path superimposed
as the thick solid line.}
\label{fig7}
\end{figure}
Note that there are pervasive pronounced oscillations due to the overlap.
This configuration is to be compared with the
approximate one obtained by adding a few terms with equal
coefficients in \cite{tucker}. Moreover, the classical path
presented in equation (33) of \cite{tucker} , describes an open
curve which seems
not to be correct. The correct parametric representation of the classical path
which is a closed curve is given by equation (\ref{eq23}).


\section{Conclusions}

We have described various classical cosmological models leading to
a class of WD equations represented by equation (\ref{eq10}). The
general solution to this equation is a superposition of products
of oscillator wave functions, as given in equation (\ref{eq19}).
We have argued that there exists a set of coefficients which
produce wave packets with the following two desired properties.
First, they posses a certain degree of smoothness, except when
there is an overlap between different branches of the wave packet,
where we expect some interference causing genuine pronounced
oscillations. Second, there is acceptable classical-quantum
correspondence. This correspondence consists of three conditions:
The wave packet should have compact support centered around the
classical path, the crest of the wave packet should follow as
closely as possible the classical path, and to each distinct
classical path there should correspond a unique wave packet with
the above properties. The set of coefficients presented here as an
ansatz is built from the coefficients of the coherent states as in
ordinary quantum mechanics and coefficients derived from the
functional form of $H_n(0)$ for even values of $n$. The resulting
wave packets can be considered as coherent states in quantum
cosmology and they posses all the aforementioned desired
properties, and the classical paths are general Lissajous figures.
We have shown these features explicitly in several examples. It is
worth mentioning that, for the class of problems discussed here,
the accomplishment of the last condition for classical-quantum
correspondence by our choice of initial conditions shows that
there is no need to employ decoherence mechanism to separate out
wave packets corresponding to distinct classical paths. However,
it is generally accepted that decoherence mechanism plays a major
role in quantum-to-classical transition, see \cite{gi} for
reviews. Although we have not shown the results explicitly here,
we can report that as $|\chi_0 |$ decreases the degree of
classical-quantum correspondence diminishes. Also note that there
is a difference between these coherent states and those of quantum
mechanics in that the former are static.

We have also shown that there
exists a continuous parameter $-\pi /2 < \theta_0 <\pi /2$, whose variation
gives a continuous interpolation between wave packets or classical paths
corresponding to different initial conditions. When $\theta_0 =0$ we
have the most symmetric configurations, {\it i.e.} loops which are symmetric
about the line $u=0$.
When $\theta_0 =\pm \pi/2 $ we
have the most asymmetric configuration, {\it i.e.} curve segments.

This work is also related to the topic of initial conditions in the context of
signature transition. For example, at the classical level there is a
controversy on the
value of $\dot{R}$ and $\dot{\phi}$ at the hypersurface of signature
transition. The authors of reference \cite{hellaby} argue that these
quantities need to be only continuous across the hypersurface, while in
\cite{hayward} it is insisted that in addition to the requirement of
continuity,
the values of both these quantities should be zero at the hypersurface.
This choice corresponds to the most asymmetric case $\theta_0 =\pm
\pi/2$, which gives curve segments as classical paths. We
speculate that these are unlikely configurations due to their pronounced
pervasive oscillations for $\omega_1/\omega_2 >1$.
\vskip20pt\noindent
{\large {\bf Acknowledgement}}\vskip5pt\noindent
The authors would like to thank the office of research of Shahid
Beheshti University for financial support.

\end{document}